\documentclass[12pt]{article}
\usepackage{latexsym,graphicx,epstopdf}
\usepackage[left=2.5cm,top=2.5cm,right=2.5cm,bottom=1.5cm]{geometry}
\usepackage{textcomp}
\usepackage{amsmath}
\usepackage{amssymb}
\usepackage{graphicx}
\begin{document}
\begin{center}
\large{\bf{FRW dark energy cosmological model with hybrid expansion law }} \\
\vspace{2mm}
\normalsize{G. K. Goswami$^1$, Anirudh Pradhan$^2$, Meena Mishra$^3$, A. Beesham$^4$ }\\
\vspace{2mm}
\normalsize{$^{1,3}$ Department of Mathematics, Kalyan P. G. College, Bhilai-490 006, C. G., India}\\
\vspace{2mm}
\normalsize{$^1$ Email: gk.goswam9@gmail.com} \\
\vspace{2mm}
\normalsize{$^3$ Email: meenamishra18@gmail.com} \\
\vspace{2mm}
\normalsize{$^2$ Department of Mathematics, Institute of Applied Sciences and Humanities, G L A University,
Mathura-281 406, Uttar Pradesh, India  \\
\vspace{2mm}
E-mail: pradhan.anirudh@gmail.com} \\
\vspace{2mm}
\normalsize{$^4$ Department of Mathematical Sciences, University of Zululand, Kwa-Dlangezwa 3886, South Africa  \\
\vspace{2mm}
E-mail: beeshama@unizulu.ac.za} \\
\end{center}
\vspace{10mm}
\begin{abstract}
In this work, we study a cosmological model of spatially homogeneous and isotropic accelerating universe which
exhibits a transition from deceleration to acceleration. For this, Friedmann Robertson Walker(FRW) metric is taken
and Hybrid expansion law $a(t)=t^{\alpha} \exp(\beta t )$ is proposed and derived. We consider the universe to
be filled with two types of fluids barotropic and dark energy which have variable equations of state. The evolution
of dark energy, Hubble, and deceleration parameters etc., have been described in the form of tables and figures. We consider
$581$ data's of observed values of distance modulus of various SNe Ia type supernovae from union $2.1$ compilation to
compare our theoretical results with observations and found that model satisfies current observational constraints.
We have also calculated the time and redshift at which acceleration in the Universe had commenced.
\end{abstract}
%%%%%%%%%%%%%%%%%%%%%%%%%%%%%%%%%%%%%%%%%%%%%%%%%%%%%%%% Section - 1 %%%%%%%%%%%%%%%%%%%%%%%%%%%%%%%%%%%%%%%
\section{Introduction}
The latest findings on  observational ground during last three decades by various
cosmological missions like observations on type Ia Supernovae (SNe Ia)
\cite{ref1}$-$\cite{ref5}, CMBR fluctuations \cite{ref6,ref7}, large scale structure
(LSS) analysis \cite{ref8,ref9}, SDSS collaboration \cite{ref10,ref11}, WMAP
collaboration \cite{ref12}, Chandra X-ray observatory \cite{ref13}, the Hubble
space telescope cluster supernova survey $V$ \cite{ref14}, BOSS
collaboration\cite{ref15}, the WiggleZ dark energy survey \cite{ref16} and latest
Planck collaboration results \cite{ref17} all confirms that our universe is
undergoing through an accelerating expansion. It is concluded that our universe
is dominated by an exotic dark energy (DE). It has negative pressure, so  is
repulsive and  creates acceleration in the universe. Inclusion of cosmological
constant in the Einstein's field equation again got importance as a positive
cosmological constant is considered as a source of Dark energy.
$\Lambda$-CMD cosmology \cite{ref18, ref19} is just Eddington-Lemaitre model
with the difference that cosmological constant term acts as a source of dark
energy with equation of state $ p_ \Lambda = \rho _ \Lambda = \frac{-\Lambda
c^4}{8\pi G}$. However, the model suffers from, inter alia, fine tuning and cosmic
coincidence problems \cite{ref20}. Any acceptable cosmological
model must be able to explain the current expansion of the universe.\\

 In any cosmological model, we require to find out rate of the expansion of the universe. Hubble constant determines it.
 Observationally we require high-precision measurement observatories to estimate Hubble constant. In general relativity,
 energy conservation equation provides linear relation ship amongst rate of the expansion, pressure, density, and temperature.
 Dark energy negative pressure and density are also included in it. There is a cosmological equation of state such as
 the relationship between temperature, pressure, and combined matter energy and vacuum energy density for any region of space.
 The problem for equation of state for barionic matter has been solved by cosmologists by providing the phases of the universe
 like stiff matter, radiation dominated and present dust dominated universe, but determination of the equation of state for
 dark energy is one of the biggest problem in observational cosmology today. Carroll and Hoffman \cite{ref21} presented a
 Dark energy (DE) model in which DE is considered in a conventional manner as a fluid by the equation of state (EoS)
 parameter $\omega_{de} = \frac{p_{de}}{\rho_{de}}$ which is not necessarily constant. We need to investigate equation
 of state (EoS) parameter for the whole span of the universe. At present, it is nearly equal to $-1$. So far two main theories
 related to variable equation of state for dark energy are quintessence and phantom models of dark energy. In quintessence model
 $-1 \leq\omega_{de}  < 0$ where as in phantom model $\omega_{de} \leq -1 $. Latest surveys such as Supernovae Legacy Survey,
 Gold sample of Hubble Space Telescope \cite{ref22,ref23}, CMB (WMAP, BOOMERANG) \cite{ref24,ref25} and large scale structure
 (Sloan Digital Sky Survey) data \cite{ref26} ruled out possibility of $\omega_{de} \ll -1$ but $\omega_{de}$ may be little
 less than $-1$. It is found that the present amount of the DE is so small as compared with the fundamental scale (fine-tuning
 problem). It is comparable with  the critical density today (coincidence problem) \cite{ref18}. So we need a different forms
 of dynamically changing DE with an effective equation of state (EoS), $\omega_{(de)} = p_{(de)}/\rho_{(de)} < -1/3$. SNe Ia
 data \cite{ref27} and a combination of SNe Ia data with CMBR anisotropy and  galaxy clustering statistics \cite{ref9} put
 limits on  $\omega_{(de)}$ as $-1.67 < \omega_{de} < -0.62$ and $-1.33 < \omega_{de} < - 0.79$, respectively.\\

 Komatsu  et al. and Hinshaw et al. \cite{ref26,ref28} estimated limits on $\omega_{(de)}$ as $-1.44 < \omega_{(de)} < -0.92$
 at $68\%$ confidence level. These observations were based on the combination of cosmological datasets coming from CMB anisotropies,
 luminosity distances of high redshift type Ia supernovae and galaxy clustering. Recently, Amirhashchi et al. \cite{ref29,ref30},
 Pradhan et al. \cite{ref31}, Saha et al. \cite{ref32}, Pradhan \cite{ref33} and  Kumar \cite{ref34} have made study on
 FRW based dark energy model in which they considered an interacting and non-interacting two type of fluids, one for
 barotropic matter and other for dark energy.\\

 In this work, we study a particular model which exhibits a transition from deceleration to acceleration. We consider
 Baryonic matter, dark energy, and ``curvature'' energy. Both baryonic matter and dark energy have variable equations of state.
 We study the evolution of the dark energy parameter within a framework of an FRW cosmological model filled with two fluids
 (barotropic and dark energy). The cosmological implications of this two-fluid scenario are discussed in detail. The model is shown
to satisfy current observational constraints.

%%%%%%%%%%%%%%%%%%%%%%%%%%%%%%%%%%%%%%%%%%%%%%%%%%%%%Section 2 %%%%%%%%%%%%%%%%%%%%%%%%%%%%%%%%%%%%%%%%%
\section{Basic field equations and their solutions}

The dynamics of the universe is governed by the Einstein's field equations (EFEs) given by

\begin{equation}
\label{1}
R_{ij}-\frac{1}{2}Rg_{ij} =-\frac{8\pi G}{c^{4}}T_{ij},
\end{equation}
where $R_{ij}$ is the Ricci tensor, $R$ is the scalar curvature, and $T_{ij}$ is the
stress-energy tensor taken as $T_{ij} = T_{ij}(m)+T_{ij}(de).$ We assume that our
universe is filled with two types of perfect fluids (since homogeneity and isotropy
imply that there is no bulk energy transport), namely an ordinary baryonic fluid
and dark energy. The energy-momentum tensors of the contents of the universe
are presented as follows: (with the subscripts $m$ and $de$ denoting ordinary
matter and dark energy, respectively).
 $T_{ij}(m)=\left(\rho_m + p_m\right)u_{i}u_{j}-p_m g_{ij}$
and $T_{ij}(de)=\left(\rho_{de}+p_{de}\right)u_{i}u_{j}-p_{de} g_{ij}$. In standard
spherical coordinates $(x^{i} = (t, r, \theta, \phi)$, a spatially homogeneous and
isotropic Friedmann-Robertson-Walker (FRW) line-element has the form (in units
$c = 1$)
\begin{equation}
\label{2}
ds{}^{2}=dt{}^{2}-a(t){}^{2}\left[\frac{dr{}^{2}}{(1+kr^{2})}+r^{2}({d\theta{}^{2}+sin{}^{2}\theta
    d\phi{}^{2}})\right],
\end{equation}
Where (i)  k=-1 is closed universe (ii)  k=1 is open universe and (iii) k=0 is
spatially flat universe. Solving EFEs (\ref{1}) for  the FRW metric (\ref{2}) and
energy momentum tensors described above, we get the following equations of
dynamic cosmology $2\frac{\ddot{a}}{a}+H^{2} = -8\pi G p + \frac{k}{a^{2}}$ and
$H^{2} =  \frac{8\pi G}{3}\rho + \frac{k}{a^{2}}$, where $H=\frac{\dot{a}}{a}$ is the
Hubble constant. Here an over dot means differentiation with respect to
cosmological time $t$. We have deliberately put the curvature term on the right
of EFEs, as this term is made to acts like an energy  term.  For this, we assume
that  the density and pressure for the curvature energy are as follows
$\rho_{k}=\frac{3 k}{8\pi Ga^{2}}, ~~ p_{k}=-\frac{k}{8\pi Ga^{2}}$. With this
choice, EFEs  are read as
 \begin{equation}\label{3}
  2\frac{\ddot{a}}{a}+H^{2} =
-8\pi G (p+p_{k}) ~\&~ H^{2}=\frac{8\pi G}{3}\,(\rho+\rho_{k}).
\end{equation}
The energy density $\rho$ in is comprised of two types of energy, namely,
matter and dark energy $\rho_m$ and $\rho_{de}$ where as the pressure `$p$' is
comprised of pressure due to matter and pressure due to dark energy. We can
express $\rho = \rho_m + \rho_{de}$, and $p =p_m+p_{de}$.

%%%%%%%%%%%%%%%%%%%%%%%%%%%%%%%%%%%%%%%%%%%%%%%%%%%% Section 3 %%%%%%%%%%%%%%%%%%%%%%%%%%%%%%%%%%%%%%%%%%%

The energy conservation law(ECL)~ $T^{ij}_{;j}=0$~ provides the following well
known equation amongst the density $\rho$, pressure $p$ and Hubble constant
$H$ $\dot{\rho}+3H(p+\rho)=0$, where  $\rho=\rho_{m}+\rho_{de}+\rho_{k}$ and
$p=p_{m}+p_{de}+p_{k},$ are the total density and pressure of the universe,
respectively. We see that $\rho_{k}$ and $p_{k}$ satisfy ECL independently, i.e.,
$\dot{\rho_{k}}+3H(p_{k}+\rho_{k})=0$, so  that
$\frac{d}{dt}{(\rho_{m}+\rho_{de})}+3H(p_{m}+p_{de}+\rho_{m}+\rho_{de})=0$. We
assume that both matter and dark energies are minimally coupled so that they
are conserved simultaneously, i.e. $\dot{\rho_{m}}+3H(p_{m}+\rho_{m})=0$, and
$\dot{\rho_{de}}+3H(p_{de}+\rho_{de})=0$. ECLs are integrable when suitable
functions are chosen relating pressure to density: $p_{m}=\omega_{m}\rho_{m}$,
$p_{de}=\omega_{de}\rho_{de}$ and $ p_{k}=\omega_{k}\rho_{k}$. In the early
part of the universe, it was radiation dominated $\omega_{m}=\frac{1}{3},\rho_{m}
\varpropto a^{-4}$. Later on when the radiation decoupled from the baryons, the
matter dominated era began. In this epoch, $ \omega_{m}=0, \rho_{m}\varpropto
a^{-3}$. In order to describe the whole history of the universe through the
equation of state, we may write $\rho_{m} = \rho_{dust} + \rho_{rad} =
(\rho_{m})_{0} \left(\frac{a_0}{a}\right)^3\left[\mu+(1-\mu)\frac{a_0}{a}\right]$,
where the parameter $\mu$ varies as per the different era of the universe. For a
particular era, it is a constant. For example, $\mu=1$ corresponds to the present
dust filled era, and $\mu=0$ corresponds to the early era of the universe in which
only radiation was present. Here terms with subscript zero are constants and
describe values at present.
 Equations of states for energies corresponding to $k$ are as follows
$ p_{k}=\omega_{k}\rho_{k}~\mbox{where}~ \omega_{k}=-1/3.$ This gives
$\rho_{k}\varpropto a^{-2}= (\rho_{k})_{0}\left[\frac{a_0}{a}\right]^2$ . We use the
relationship between scale factor and red shift as $\frac{a_0}{a} = 1+z$,
 to write variables in terms of redshift $z$ rather than time.
 The critical density and the density parameters for energy density, dark
energy and curvature density (in units $c = 1$) are, respectively, defined by $
\rho_{c}=\frac{3H^{2}}{8\pi G}$, $\Omega_{m}=\frac{\rho_{m}}{\rho_{c}}$, $
\Omega_{de}=\frac{\rho_{de}}{\rho_{c}}$, and
$\Omega_{k}=\frac{\rho_{k}}{\rho_{c}}$.
 With these in hand, we can write the FRW field equations as follows
  \begin{equation}
  \label{4}
   H^2=H^{2}_{0}\left[(\Omega_{m})_{0} \left(\frac{a_0}{a}\right)^{3}\left[\mu+(1-\mu)\frac{a_0}{a}\right]+(\Omega_{k})_{0}
   \left(\frac{a_0}{a}\right)^{2} \right] + H^2 \Omega_{de},
  \end{equation}
  and
  \begin{equation}
  \label{5}
   2q = 1 + \frac{3H^2_0}{H^2}\left[\omega_{m}(\Omega_{m})_{0} \left(\frac{a_0}{a}\right)^{3}\left[\mu+(1-\mu)\frac{a_0}{a}\right]
   - \frac{1}{3}(\Omega_{k})_{0} \left(\frac{a_0}{a}\right)^2\right]+ 3\omega_{de}\Omega_{de} ,
  \end{equation}
  where $q$ is the deceleration parameter defined by $ q=-\frac{\ddot{a}}{aH^2}$.
It is important to mention here that this model represents a decelerating universe
in the absence of dark energy. This is so because the deceleration parameter
$q$ is positive when the dark energy is zero. We recall  $q\lesseqqgtr 1/2$ for
closed, flat and open universes, respectively. The $\Lambda$ CMD  model fits
best with the present day observations. In this model, $\Lambda$ accounts for
the vacuum energy with its energy density $\rho_{\Lambda}$ and pressure
$p_{\Lambda}$ satisfying the equation of state (EoS) $ p_{\Lambda} = -
\rho_{\Lambda} = \frac{\Lambda}{8\pi G}; ~ ~ \omega_{de} = -1$. The purpose of
this paper is   to investigate the  effect of  $\omega_{de}$  as a  function of time.

%%%%%%%%%%%%%%%%%%%%%%%%%%%%%%%%%%%%%%%%%%%%%%%%%%%%%%%%%%% Section 4 %%%%%%%%%%%%%%%%%%%%%%%%%%%%%%%%%%%%%%%%%%%%%%%%

We have only two equations and the scale factor `$a$', pressure $p$ and energy density $\rho$ to be determined. So we have to
use a certain ansatz. In e print \cite{ref35}, we have developed hybrid expansion law [HEL] for scale factor 'a' as follows  
 $a(t)=t^{\alpha} \exp(\beta t )$, where $\alpha$ and $\beta$ are arbitrary constants. These have been obtained as
 $ \beta = 0.0397474\sim 0.04, \alpha = 0.415066 \sim 0.415$ on the basis of Planck's observational results  \cite{ref17} 
 and the following differential equations obtained from HEL.
  \begin{equation}
 \label{6}
 \alpha (1+z) H H_z =  \alpha (q+1)  H^2 = (H-\beta)^2 = \frac{\alpha^2}{t^2}.
 \end{equation}
 
  %%%%%%%%%%%%%%%%%%%%%%%%%%%%%%%%%%%%%%%%%%%%%%%%%%%%% Subsection 4.1 %%%%%%%%%%%%%%%%%%%%%%%%%%%%%%%%%%%%%%%%%
  \subsection{Hubble constant $H$ }
 
  Eq. (\ref{6}) is solved for the Hubble constant $H$ as a function of redshift $z$ as follows

  \begin{equation}
  \label{7}
  (H-\beta)^{\alpha}= A~\exp\left(\frac{\alpha\beta}{H-\beta}\right) (1+z),
  \end{equation}
  where constant of integration A is obtained as $A = 0.134$ on the basis of the present value of $H (H_0=0.07$ Gy$^{-1})$.
   A numerical solution of Eq. (\ref{7}) shows that the Hubble constant is an increasing function of red shift. \\

    %%%%%%%%%%%%%%%%%%%%%%%%%%%%%%%%%Figure 1 %%%%%%%%%%%%%%%%%%%%%%%%%%%%%%%%%%%%%%%%%%%%%%%%%%%%%
  %\begin{figure}[!h]
  \begin{figure}[ht]
\centering
    \includegraphics[width=10cm,height=5cm,angle=0]{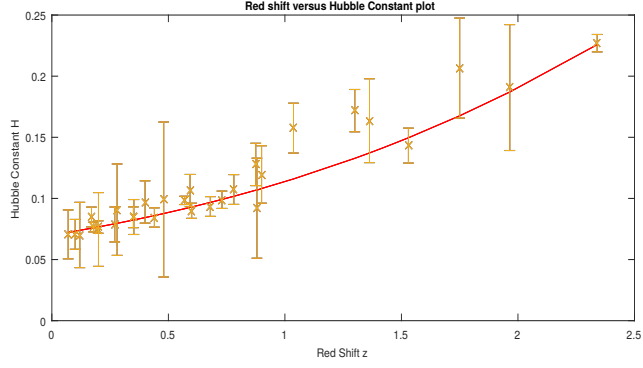}
    \caption{Plot of Hubble constant ($H$) versus redshift ($z$)}
  \end{figure}
  %%%%%%%%%%%%%%%%%%%%%%%%%%%%%%%%%%%%%%%%%%%%%%%%%%%%%%%%%%%%%%%%%%%%%%%%%%%%%%%%%%%%
  %%%%%%%%%%%%%%%%%%%%%%%%%%%%%%%%%%%%%% Figure 2 %%%%%%%%%%%%%%%%%%%%%%%%%%%%%%%%%%
%\begin{figure}[!h]
\begin{figure}[ht]
\centering
    \includegraphics[width=10cm,height=5cm,angle=0]{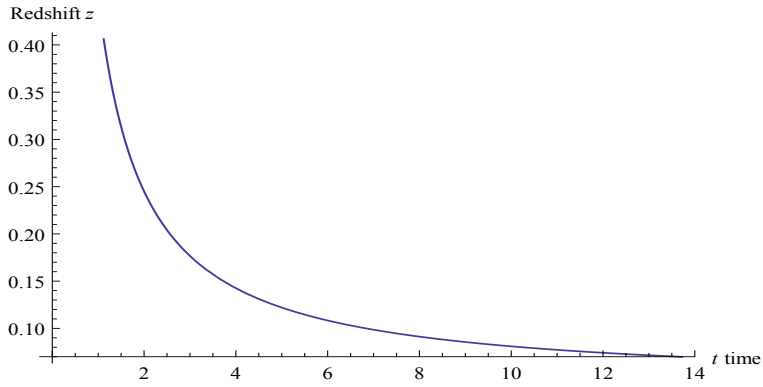}
    \caption{Plot Redshift ($z$) versus time ($t$) }
\end{figure}
%%%%%%%%%%%%%%%%%%%%%%%%%%%%%%%%%%%%%%%%%%%%%%%%%%%%%%%%%Subsection 4.2 %%%%%%%%%%%%%%%%%%%%%%%%%%%%%%%%%%%%%%%%%%%%

 As it is clear that Hubble constant is in fact not a constant. It varies slowly over red shift and time. Various researchers
 \cite{ref15,ref16,ref36,ref37,ref38,ref39} have estimated values of Hubble constant at different red-shifts using Differential
age approach and Galaxy Clustering method. The following table describes various observed values of Hubble constant Hob along
with corrections as per red-shift in the range $0\leq z \leq 2$. We have also calculated and put the corresponding theoretical
values of $H$ as per our model in the table. It is found that both observed and theoretical values tally considerably and support
our model.

\begin{center}
[Table-1]\\(The values of the Hubble constant ($H$) for redshift ($z$) ranging  between $0$ to $2$ )\\
(Hubble constant is expressed in Gy $^{-1}$ )

\end{center}
$\begin{array}{|c|c|c|c|c|c|c|c|}
\hline
z &  0.07 & 0.1 & 0.12 & 0.17 & 0.179 & 0.199 \\
\hline
Hob & .0706 \pm .0200  &   .0706  \pm   .0122 & .0701 \pm .0267 & .0848 \pm .0081 & .0766  \pm  .0040  & .0766 \pm .0051 \\
\hline
Hth &  0.0722454 & 0.0732392  & 0.0739125  & 0.0756344  & 0.0759503  & 0.0766589\\
\hline
\end{array}$
\\
\\
$\begin{array}{|c|c|c|c|c|c|c|c|}
\hline
z & 0.2  & 0.27 & 0.28 & 0.35 & 0.352 & 0.4  \\
\hline
Hob &  .0746 \pm  .0302 & .0787  \pm .0143 & .0908 \pm  .0374 & .0846 \pm .0085  & .0848 \pm .0143  & .0971 \pm .0173 \\
\hline
Hth & 0.0766946 & 0.0792498 & 0.0796243 & 0.0823145 & 0.0823932 & 0.0843112\\
\hline
\end{array}$ \\

$\begin{array}{|c|c|c|c|c|c|c|c|}
\hline
z & 0.44  & 0.48 & 0.57 & 0.593 & 0.60 & 0.68 \\
\hline
H0b &  .0844 \pm .0079 & .0991  \pm.0634 &  .0984  \pm .0034  & .1064 \pm .0132 &  .0899 \pm .0062  & .0934 \pm .0081  \\
\hline
Hth & 0.0859548 & 0.0876406 & 0.0915919 & 0.0926377 & 0.092959 & 0.0967302\\
\hline
\end{array}$ \\

$\begin{array}{|c|c|c|c|c|c|c|c|}
\hline
z &    0.73 & 0.781 & 0.875 & 0.88 & 0.90 &     1.037   \\
\hline
Hob &   .0989 \pm .0071 & 0.1073 \pm .0122 & 0.1278 \pm .0173 & .0920 \pm .0409 & .1196 \pm .0235 & .1575 \pm .0204 \\
\hline
Hth & 0.0991824 & 0.101761 & 0.106723 & 0.106723 & 0.10809 & 0.115938 \\
\hline
\end{array}$\\
\\
\\
$\begin{array}{|c|c|c|c|c|c|c|c|}
\hline
z &1.3 & 1.363 &    1.43 &  1.53 &  1.75 & 1.965 \\
\hline
Hob & .1718 \pm .0173 & .1636 \pm .0343 & .1810 \pm .0184 & .1432 \pm .0143 & .2066 \pm .0409 & .1907 \pm .0515  \\
\hline
Hth & 0.132796 & 0.137201 & 0.142047 & 0.149595 & 0.167577 & 0.187053 \\
\hline
\end{array}$\\
\\
\\
Figure $1$ depicts the variation of the Hubble constant $H$ with red shift $z$.
From this figure, we observe that $H$ increases with the increase of red shift. In
this figure, cross signs are $31$ observed values of Hubble constants $H_{0}$
with corrections where as the linear curve is the theoretical graph of the Hubble
constant $H$ as per our model. Figure $2$ plots the variation of red shift $z$
with time $t$ which shows that in the early universe the red shift was more than
at present.

  %%%%%%%%%%%%%%%%%%%%%%%%%%%%%%%%%%%%%%%%%%%%%% Subsection 4.2 %%%%%%%%%%%%%%%%%%%%%%%%%%%%%%%%%%%%%%%%%%%
  \subsection{ DE parameter $\Omega_{de}$ and equation of state parameter $\omega_{de}$ for DE density}

  Now, from Eqs.(\ref{4}), (\ref{5}) and (\ref{6}), the density parameter $\Omega_{de}$ and
  EoS parameter $\omega_{de}$ for dark energy are given by
 \begin{equation}
\label{8}
H^2 \Omega_{de} = H^2 - (\Omega_{m})_0 H^2_0 (1+z)^3 -(\Omega_{k})_0 H^2_0 (1+z)^2
\end{equation}

\begin{equation}
\label{9}
\omega_{de}=\frac{(2-3\alpha)H^2-4\beta H + 2 \beta^2 + \alpha (\Omega_{k})_0 H^2_0 (1+z)^2 }
{3 \alpha [H^2-H_0^2(\Omega_{m})_0(1+z)^{3}-H_0^2(\Omega_{k})_0(1+z)^{2}]}.
\end{equation}
Where we have taken $\omega_{m}=0$ and $\mu=1$ for present dust filled universe.

  We present the following numerical tables [$2$] and [$3$], which display values of the energy parameter $\Omega_{de}$ and
  EoS parameter $\omega_{de}$ for DE versus red shift $z$ ranging between $0$ to $1.4$.

   \begin{center}
    [Table-$2$]\\(The values of $\Omega_{de}$ and $z$ ranging between $0$ to $3$ )
  \end{center}

  $\begin{array}{|c|c|c|c|c|c|c|c|c|}
  \hline
  z &  0.0  & 0.2  & 0.4  & 0.6  & 0.8 &  1.0   \\
  \hline
   \Omega_{de} &  0.695 &  0.671884 &  0.609804 &  0.506256  & 0.360992 &  0.176341 \\
  \hline
  \end{array}$\\

  \begin{center}
    [Table-3]\\(The values of EoS parameter for DE ($\omega_{de}$) and red shift ($z$) ranging
   between $0$ to $1$ )
  \end{center}

  $\begin{array}{|c|c|c|c|c|c|}
  \hline
   z &  0. & 0.1 & 0.2 & 0.3 & 0.4   \\
  \hline
  \omega_{de} & -1.00673 & -0.961887 & -0.926735 & -0.90137 & -0.886524  \\
  \hline
  \end{array}$\\
  \vspace{5mm}
  $\begin{array}{|c|c|c|c|c|c|c|}
  \hline
   z  & 0.5 & 0.6 & 0.7 & 0.8 & 0.9 & 1.0  \\
  \hline
   \omega_{de} & -0.883867 & -0.896634 & -0.930997 & -0.999402 & -1.13007 & -1.40182  \\
  \hline
  \end{array}$\\
  \\
  \\
 Figure $3$ depicts the variation of $\Omega_{de}$  with $z$. Figure $4$ plots the variation of $\omega_{de}$ with $z$.
 These Figures $3$ and $4$ support the content of the tables $2$ and $3$. \\

 %%%%%%%%%%%%%%%%%%%%%%%%%%%%%%%%%%%%%%%%%%%%%%%%% Figure 3 %%%%%%%%%%%%%%%%%%%%%%%%%%%
   %\begin{figure}[!h]
   \begin{figure}[ht]
\centering
    \includegraphics[width=10cm,height=5cm,angle=0]{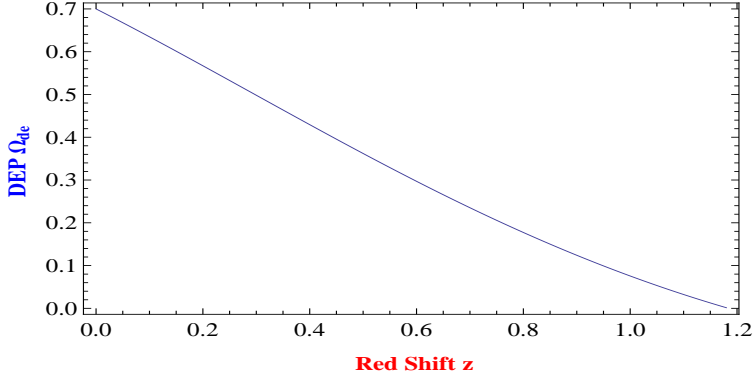}
    \caption{Plot of Energy density parameter for DE ($\Omega_{de}$) versus redshift ($z$)}
  \end{figure}
  %%%%%%%%%%%%%%%%%%%%%%%%%%%%%%%%%%%%%%%%%%%%%%%%%%%%%%%%%%%%%%%%%%%%%%%%%%%%%%%%%%%%%%%%%%
  %%%%%%%%%%%%%%%%%%%%%%%%%%%%%%%%%%%%%% Figure 4 %%%%%%%%%%%%%%%%%%%%%%%%%%%%%%%%%%%%%%
  %\begin{figure}[!h]
  \begin{figure}[ht]
\centering
    \includegraphics[width=10cm,height=5cm,angle=0]{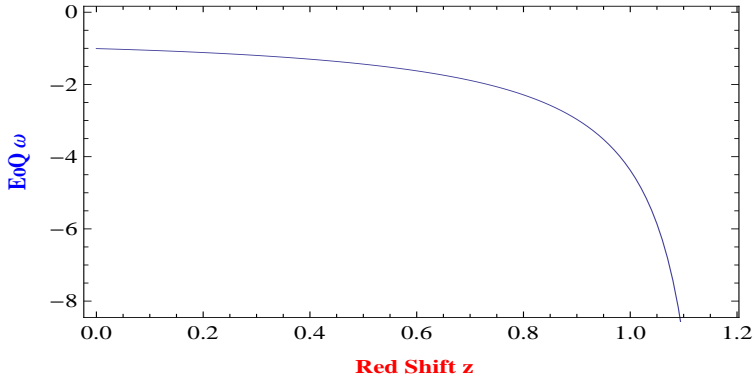}
    \caption{EoS parameter for DE ($\omega_{de}$) versus redshift ($z$)}
  \end{figure}
  %%%%%%%%%%%%%%%%%%%%%%%%%%%%%%%%%%%%%%%%%%%%%%%%%%%%%%%%%%%%%%%%%%%%%%%%%%%%%%%%%%%%%%%%%%%%
It is found that at $z \simeq 1.16321$, there is a singularity while determining $ \omega_{de}$. So we can say
that this model works well during  $ 0 \leq z \leq 1.1632$. Moreover, $ -1.68889\leq \omega_{de} \leq -0.910382 $
during  $ 0 \leq z \leq 0.98$. This is the limit on $\omega_{de}$ found by various surveys. The recent
supernovae SNI $997ff$ at $z \simeq 1.7$ is consistent with a decelerated expansion at the epoch of high
emission ( Benitez et al. \cite{ref40}, Turner \& Riess \cite{ref41}).

%%%%%%%%%%%%%%%%%%%%%%%%%%%%%%%%%%%%%%%%%%%%%%%%%%%%% Subsection 4.3 %%%%%%%%%%%%%%%%%%%%%%%%%%%%%%%%%%%%%%
 \subsection{ Time at which energy density parameter for DE $\Omega_{de}$ opposes deceleration }

From Eqs. (\ref{7}) and (\ref{8}), we observe that \\

$z \to 1.1633,~~\Omega_{de}\to -0.0000984892$ and when $z \to 1.1632 ,~~\Omega_{de}\to 0.0000157204.$ \\

It means \\

$\Omega_{de}\to 0$, when $z \to 1.16325~~$  i.e.~~ $t \to 4.91961 ~ ~ Gyrs. $\\

At this time $ q \to 0.092543 $. Thus, as per our model, dark energy begins its role of opposing deceleration and exerting
 negative pressure at time  $t \to 4.91961 ~ ~ Gyrs$.

%%%%%%%%%%%%%%%%%%%%%%%%%%%%%%%%%%%%%%%%%%%% Subsection 4.4%%%%%%%%%%%%%%%%%%%%%%%%%%%%%%%%%%%%%%%%%%%%
\subsection{ Densities in our model}
From ECLs for the matter density  and dark energy density  we obtain:
$\rho_{m}= (\Omega_{m})_0(1+z)^3(\mu+(1-\mu)(1+z))(\rho_{c})_0$ and $
\rho_{de}= \Omega_{de}[(\Omega_{m})_0
[1+z]^3[\mu+(1-\mu)(1+z)]+(\Omega_{k})_0  [1+z]^2+ \Omega_{de}](\rho_{c})_0,$
where $(\rho_{c})_0 = 1.88\times10^{-29}gm/cm^{3}.$
 Now we present the following table which describes the dark energy density $\rho_{de}$ in units  of $(\rho_{c})_0$
 for redshift $z$ ranging between $0$ and $1$.
 \begin{center}
    [Table-4]\\(The values of energy density for DE ($\rho_{de}$) for redshift ($z$) ranging in between $0$  and $1$ )
\end{center}
$\begin{array}{|c|c|c|c|c|c|}
\hline
 z &    0.0 & 0.1 & 0.2 & 0.3 & 0.4  \\
\hline
 \rho_{de}/(\rho_{c})_0 & 0.695 & 0.752362 & 0.80457 & 0.848396 & 0.879828  \\
\hline
\end{array}$\\\\
$\begin{array}{|c|c|c|c|c|c|c|}
\hline
 z & 0.5 & 0.6 & 0.7 & 0.8 & 0.9 & 1.0 \\
\hline
\rho_{de}/(\rho_{c})_0 & 0.893955 & 0.884863 & 0.845532 & 0.767754 & 0.642085 & 0.45784 \\
\hline
\end{array}$
  %%%%%%%%%%%%%%%%%%%%%%%%%%%%%%%%%%%%%%%%%%%%%%%%%%%%%%% Subsection  4.5 %%%%%%%%%%%%%%%%%%%%%%%%%%%%%%%%%%%%%%%
  \subsection{ Transition from deceleration to acceleration }
 Now we present the following table which describes the transition from deceleration to acceleration.
 \begin{center}
   [Table-$5$]\\(The values of the deceleration parameter ($q$) for redshift ($z$) ranging between $0$  and $10$ )
        \end{center}
  $\begin{array}{|c|c|c|c|c|c|}
  \hline
   z &  0 & 1 & 2 & 3 & 4  \\
  \hline
   q & -0.552016 & 0.0253422 & 0.521904 & 0.849706 & 1.04866   \\
  \hline
  \end{array}$

  $\begin{array}{|c|c|c|c|c|c|c|}
  \hline
   z & 5 & 6 & 7 & 8 & 9 & 10 \\
  \hline
   q     & 1.16973 & 1.24585 & 1.29563 & 1.32943 & 1.35318 & 1.37036\\
  \hline
  \end{array}$

  %%%%%%%%%%%%%%%%%%%%%%%%%%%%%%%%%%%%%%%%%% Figure 5 %%%%%%%%%%%%%%%%%%%%%%%%%%%%%%
 \begin{figure}[ht]
\centering
    \includegraphics[width=10cm,height=5cm,angle=0]{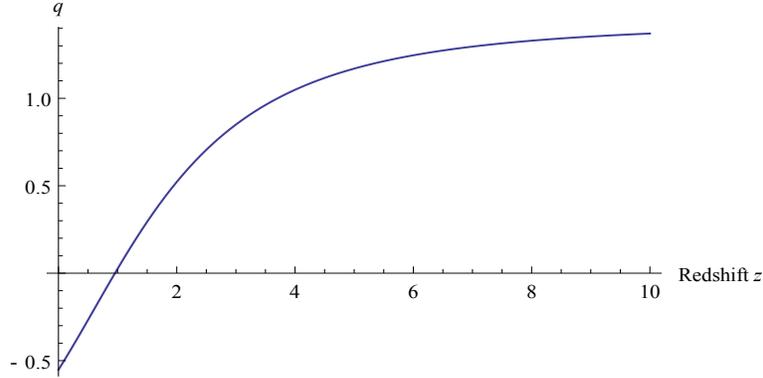}
    \caption{Deceleration parameter $q$ versus red shift $z$}
  \end{figure}
  %%%%%%%%%%%%%%%%%%%%%%%%%%%%%%%%%%%%%%%%%%%%%%%%%%%%%%%%%%%%%%%%%%%%%%%%%%%%%%%%%%%%

  Figure $5$ depicts the variation of DP ($q$) with red shift $z$ based on  the above table. The figure describes the fact more clearly.

  %%%%%%%%%%%%%%%%%%%%%%%%%%%%%%%%%%%%%%%%%%%%%%%%%%%%%%% Subsection 4.6 %%%%%%%%%%%%%%%%%%%%%%%%%%%%%%%%%%%%%%%%%
  \subsection{Time at which acceleration had began}
  At $z=0.9557,~ \&~ 0.9558$, our model gives following values of Hubble constant $H$, deceleration  parameter $q$ and
  and corresponding time.\\

 $$H(0.9557)\to 0.111206,~ ~ q(0.9557)\to -0.0000124355, ~ ~t(0.9557)\to 5.81124 ,$$
 and
  $$H(0.9558)\to 0.111212, ~ ~ q(0.9558)\to 0.0000450098, ~ ~ t(0.9558)\to 5.81078 .$$

 This means that the acceleration had begun at $ z\to 0.95575,t\to 5.81104~ Gyr, H\to 0.111209 ~Gyr^{-1} $. At this time
 $ \Omega_{de}=0.220369 $ and $\omega_{de}=-1.54715$.

  %%%%%%%%%%%%%%%%%%%%%%%%%%%%%%%%%%%%%%%%%%%%%%%%%%%%%% Section 5 %%%%%%%%%%%%%%%%%%%%%%%%%%%%%%%%%%%%%%%%%%%%%5
  \section{Luminosity distance versus red shift relation}

 Red shift-luminosity distance relation \cite{ref21,ref42} ia an important observational tool to study
 the evolution of the universe. The expression for luminosity distance($D_L$) is obtained in term of red shift as the light coming out
 of a distant luminous body get red shifted due to the expansion of the universe. We determine the flux of a source with the help of
 luminosity distance. It is given as $D_{L}=a_{0} r (1+z)$, where r is the radial co ordinate of the source.
  We consider a ray of light having initially  $ \frac{d\theta}{ds}=0 ~ ~ \mbox{and} ~ ~ \frac{d\phi}{ds}=0$,
  then  geodesic for the metric (\ref{5}) will determine $ \frac{d^2\theta}{ds^2}=0 ~ ~ \mbox{and} ~ ~ \frac{d^2\phi}{ds^2}=0$.
So if we pick up a light ray in a radial direction, then it continues to move along
the $r$-direction always, and we get following equation for the path of light
$ds^{2}= c^{2}dt^{2}- \frac{a^{2}}{1+kr^2}dr^2=0$. As we have seen, the effect of
curvature is very small at present, $(\Omega_{k})_0=0.005$, so for the sake of
  simplification, we take $k=0$. From this we obtain
  %%%%%%%%%%%%%%%%%%%%%%%%%%%%%%%%%%%%%%%%%%%%%%%% Figure 6 %%%%%%%%%%%%%%%%%%%%%%%%%%%%
%\begin{figure}[!h]
\begin{figure}[ht]
\centering
    \includegraphics[width=10cm,height=5cm,angle=0]{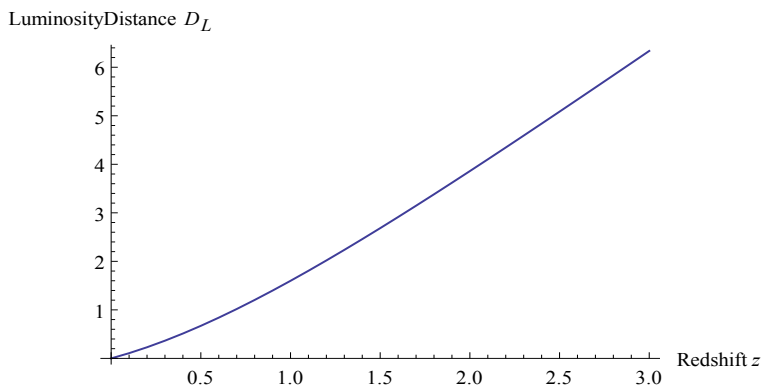}
    \caption{ Luminosity distance ($D_{L}$) $-$ Redshift ($z$) Plot}
\end{figure}
  %%%%%%%%%%%%%%%%%%%%%%%%%%%%%%%%%%%%%%%%%%%%%%%%%%%%%%%%%%%%%%%%%%%%%%%%%

    $r  =  \int^r_{0}dr = \int^t_{0}\frac{cdt}{a(t)} = \frac{1}{a_{0}H_{0}}\int^z_0\frac{cdz}{h(z)}$,
where we have used $ dt=dz/\dot{z}$,  $\dot{z}=-H(1+z)$ and
$h(z)=\frac{H}{H_0}$. So we get following expression for the luminosity distance
  \begin{equation}\label{10}
 D_{L}=\frac{c(1+z)}{H_{0}}\int^z_0\frac{dz}{h(z)}
 \end{equation}
 In our earlier work \cite{ref43,ref44}, we already obtained luminosity distance.\\

Solving Eqs. (\ref{7}) $-$ (\ref{9})  and   (\ref{10})  numerically, we get following
table for values of luminosity distance ($D_{L}$) at various red shifts.

 \begin{center}
  [Table-6]\\( Luminosity distances ($D_{L}$) at redshifts ($z$) in the range $0$ to $3$ )
 \end{center}

 $ \begin{array}{|c|c|c|c|c|c|c|c|c|}
  \hline
z & 0.0 &      0.2  &    0.4 &       0.6  &      0.8  &    1.0  &      1.2  & 1.4    \\
 \hline
  D_{L} & 0.0 &   0.229565  & 0.512198 &  0.839454 &  1.2038  & 1.59863 &  2.01827 & 2.4579    \\
  \hline
\end{array} $ \\
  $ \begin{array}{|c|c|c|c|c|c|c|c|c|}
   \hline
 z &       1.6  &     1.8  &     2.0      &       2.2 &      2.4 &     2.6  &    2.8    &    3.0 \\
  \hline
  D_{L} & 2.9135   & 3.38175  & 3.85993 &     4.34584 &  4.83769  &  5.33408     &  5.83385  &   6.33612   \\
  \hline
\end{array} $

%%%%%%%%%%%%%%%%%%%%%%%%%%%%%%%%%%%%%%%%%%%%%%%%%%%%%%%%%%%%%%%%%%%%%%%%%%%%%%%%%%%%
\vspace{5mm}

Figure $6$ depicts the variation of the luminosity distance with red shift based on
Table-$6$. We observe that the luminosity distance  is an increasing function of
red shift.
%%%%%%%%%%%%%%%%%%%%%%%%%%%%%%%%%%%%%%%%% Section 6 %%%%%%%%%%%%%%%%%%%%%%%%%%%%%%%%%

 \section{ Distance modulus $\mu$ and apparent magnitude $m_{b}$ for type Ia Supernovas (SNe Ia):}

The distance modulus $\mu$ of a source is defined as $ \mu = m_{b}-M$, where
$m_{b}$ and $M$ are the apparent and absolute magnitude of the source,
respectively.  The distance  modulus is related to the luminosity distance
through the following formula $ \mu = M-m_{b} = 5 log_{10}
\left(\frac{D_L}{Mpc}\right) + 25 $. Type Ia supernova (SN Ia) are standard candle.
They have a common absolute magnitude $M$ irrespective of the red shift $z$.
 We use following equation for luminosity distance for a supernova at very small red shift
  $D_L=\frac{cz}{H_0}$.

  %%%%%%%%%%%%%%%%%%%%%%%%%%%%%%%%%%% Figure 7 %%%%%%%%%%%%%%%%%%%%%%%%%%%%%%%%%%%%%%%%%%%%%%%%%%%%%%%
%\begin{figure}[!h]
\begin{figure}[!ht]
    \includegraphics{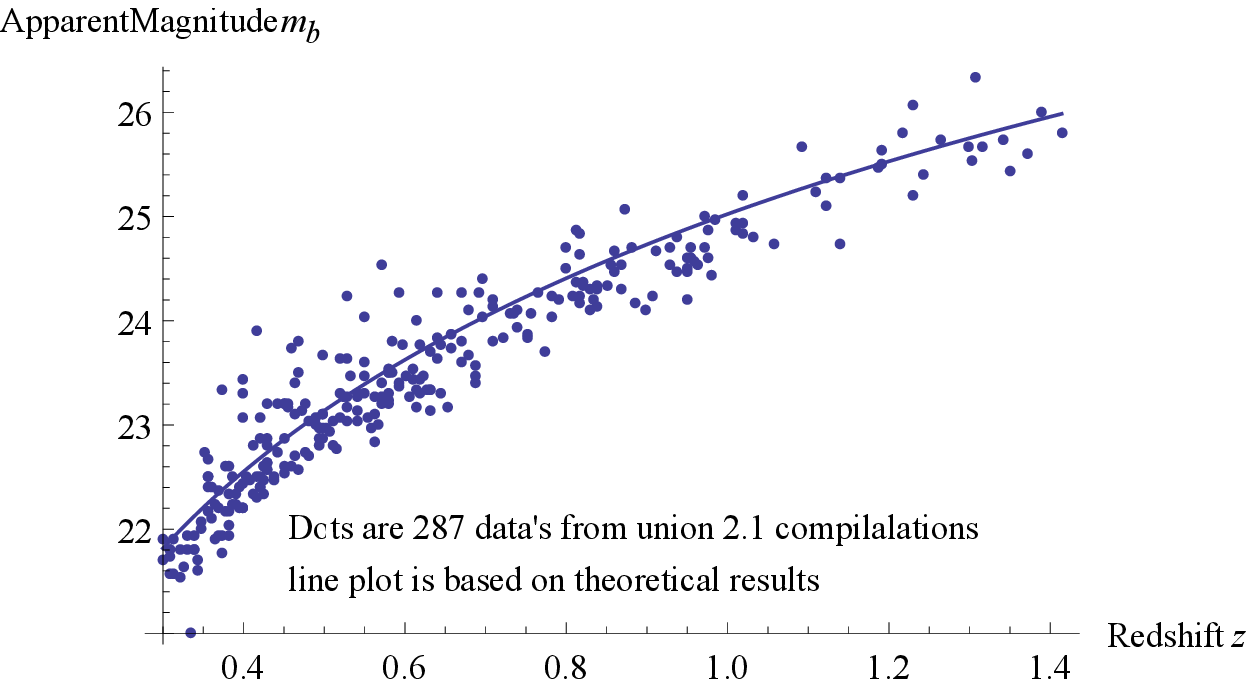}
    \caption{Plot of apparent magnitude ($m_b$) versus red shift ($z$)}
\end{figure}
%%%%%%%%%%%%%%%%%%%%%%%%%%%%%%%%%%%%%%%%%%%%%%%%%%%%%%%%%%%%%%%%%%%%%%%%%%%%%%%
%%%%%%%%%%%%%%%%%%%%%%%%%%%%%%%%%%%%% Figure 8 %%%%%%%%%%%%%%%%%%%%%%%%%%%%%%%%%%%%%%%%%%%
%\begin{figure}
\begin{figure}[!ht]
    \includegraphics{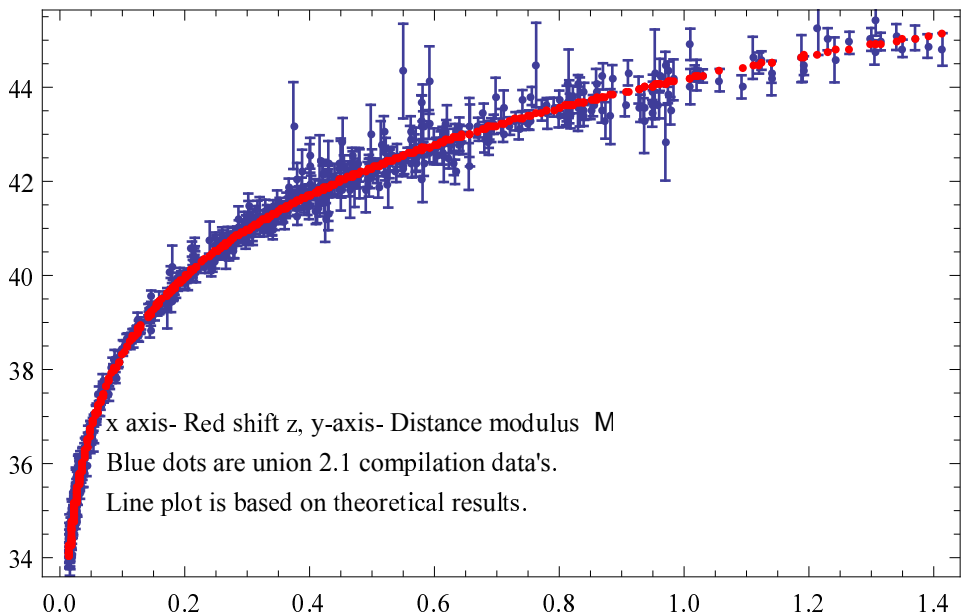}
    \caption{Plot of distance modulus ($\mu=M-m_b$) versus redshift ($z$)}
\end{figure}
%%%%%%%%%%%%%%%%%%%%%%%%%%%%%%%%%%%%%%%%%%%%%%%%%%%%%%%%%%%%%%%%%%%%%%%%%%%%%%%%%%%%%%%%%%%%%%%%%
  %%%%%%%%%%%%%%%%%%%%%%%%%%%%%%%%%%%%%%%%%%%%%%%%%%%%%%%%%%%%%%%%%%%%%%%%%%%%%%%%%%%%%%%%%
   In the literature there are so many supernova of low red shift whose apparent magnitudes are known. These determines
   common absolute magnitude  $M$ of all SN Ia supernovae.
   In our earlier work \cite{ref43,ref44}, we have obtained $M$ as follows $  M = 5log_{10}\left(\frac{H_0}{.026 c}\right)- 8.92$.
   From these, we obtain, $log_{10}(H_{0}D_{L})= (m_{b} - 16.08)/5 + log_{10}(.026c)$ and following expression for apparent magnitudes
   \begin{equation}\label{12}
 m_{b} = 16.08+ 5log_{10}\left[\frac{1+z}{.026} \int^z_0\frac{dz}{h(z)}\right].
 \end{equation}
  Solving Eqs. (\ref{7}) $-$ (\ref{9}) and (\ref{12}) numerically, we present  the following table describing
  the values of the apparent magnitude $m_{b}$ for redshift $z$ ranging between $0$ and $1.6$
  \begin{center}
  [Table-7]\\(The values of apparent magnitude ($m_{b}$) for redshift ($z$) in the range $0$ to $1.6$ )
  \end{center}
  $ \begin{array}{|c|c|c|c|c|c|c|c|c|}
  \hline
  z &   0.1 & 0.2 & 0.3 & 0.4 & 0.5 & 0.6 & 0.7 & 0.8 \\
  \hline
  m_{b} &   19.1639 & 20.8097 & 21.8154 & 22.5523 & 23.1379 & 23.6251 & 24.0426 & 24.4079    \\
  \hline
  \end{array} $ \\
  $ \begin{array}{|c|c|c|c|c|c|c|c|c|}
  \hline
  z &   0.9 & 1. & 1.1 & 1.2 & 1.3 & 1.4 & 1.5 & 1.6     \\
  \hline
  m_{b}  & 24.7323 & 25.0239 & 25.2883 & 25.53 & 25.7523 & 25.958 & 26.149 & 26.3272 \\
  \hline
  \end{array} $
  \\
  \\
  We consider $581$ data points of the observed values of the distance modulus of various SNe Ia type supernovas from the union $2.1$
  compilation \cite{ref14} with red shift in the range $z\leq 1.414$. We calculate  the  corresponding theoretical values
  as per our model. The following Figures $7$ \& $8$ depict the closeness of observational and theoretical results, thereby justifying
   our model.

%%%%%%%%%%%%%%%%%%%%%%%%%%%%%%%%%%%%%%%%%% Section 7 %%%%%%%%%%%%%%%%%%%%%%%%%%%%%%%%%%%%%%%%%%
 \section{ Conclusions}
In this work, efforts are made to develop a cosmological model which satisfies the cosmological
principle and incorporates the latest development which envisaged that our universe is accelerating due to dark energy.
We have also proposed a variable equation of state for dark energy in our model. We studied a model with radiation, dust
and dark energy which shows a transition from deceleration to acceleration. We have successfully subjected our model to
various observational tests.

In a nutshell, we believe that our study will pave the way to more research in future, in particular, in the area of the early
universe, inflation and galaxy formation, etc. The proposed hybrid expansion law may help in the investigations of
hidden matter like dark matter, dark energy and black holes.

\section*{Acknowledgement} The authors (G. K. Goswami \& A. Pradhan) sincerely acknowledge the Inter-University Centre for Astronomy
and Astrophysics (IUCAA), Pune, India for providing facility where part of this
work was completed during a visit. A. Pradhan would also like to thank the
University of Zululand, South Africa for providing facilities and support where part
of this work has been done. The authors would like to convey their sincere
thanks to Sergei D. Odintsov ,Kazuharu Bamba  and Sunny Vagnozzi for  useful
suggestions and providing references \cite{ref45}$-$\cite{ref52} for the
improvement of the paper in present form.

%%%%%%%%%%%%%%%%%%%%%%%%%%%%%%%%%%%%%%%%%%%%%%%%%%%%%%%%%%%%%%%%%%%%%
%\newpage

  \end{document}